# Multiple Andreev reflections in topological insulator nanoribbons

Rak-Hee Kim[1], Nam-Hee Kim[1], Bongkeon Kim[1], Yasen Hou[2], Dong Yu[2] and Yong-Joo Doh[1]

[1]Department of Physics and Photon Science, Gwangju Institute of Science and Technology (GIST), Gwangju 61005, Korea
[2]Department of Physics, University of California, Davis, California 95616, United States

**Corresponding Authors**

[*]E-mail: yjdoh@gist.ac.kr

**Abstract**

Superconducting proximity junctions made of topological insulator (TI) nanoribbons (NRs) provide a useful platform for studying topological superconductivity. We report on the fabrication and measurement of Josephson junctions (JJs) using Sb-doped $Bi_2Se_3$ NRs in contact with Al electrodes. Aharonov–Bohm and Altshuler–Aronov–Spivak oscillations of the axial magneto-conductance of TI NR were observed, indicating the existence of metallic surface states along the circumference of the TI NR. We observed the supercurrent in the TI NR JJ and subharmonic gap structures of the differential conductance due to multiple Andreev reflections. The interface transparency of the TI NR JJs estimated based on the excess current reaches $\tau = 0.83$, which is among the highest values reported for TI JJs. The temperature dependence of critical current is consistent with the short and ballistic junction model confirming the formation of highly transparent superconducting contacts on the TI NR. Our observations would be useful for exploring topological Josephson effects in TI NRs.

## 1. Introduction

Three-dimensional topological insulators (TIs) are bulk insulators containing metallic surface states topologically protected by time-reversal symmetry [1]. The strong spin–orbit coupling in TI induces the spin orientation of the surface electron to be locked perpendicular to its translational momentum, the so-called spin–momentum locking [2]. Especially, the topological surface states in TI nanoribbons (NRs) are modified to form discrete one-dimensional (1D) subbands, caused by the quantum confinement effect along the circumference of the TI NR [3]. Applied with the axial magnetic field $B_{\text{axial}}$, TI NRs exhibit periodic oscillations of magneto-conductance (MC) with a period of the flux quantum given by $\Phi_0 = h/e$, where $h$ is the Planck constant and $e$ is the elementary charge, owing to the Aharonov–Bohm (AB) effect [4] combined with the Berry phase $\pi$ [3]. The AB oscillations were observed in various NRs made of $Bi_2Se_3$ [5, 6], $Bi_2Te_3$ [7, 8], and Dirac semimetals [8, 9]. Moreover, Altshuler–Aronov–Spivak (AAS) oscillations ($h/2e$-periodic oscillations) [10] occur concurrently with the AB oscillations because of the quantum interference between a pair of time-reversed paths along the circumference of the TI NR [11].

When the TI NR forms a superconducting contact with the conventional superconductor [3, 12], it is theoretically expected that there forms a zero-gap 1D mode hosting the Majorana bound states (MBS) [13], which provide a building block for developing fault-tolerant topological quantum computation [14]. The existence of the MBS can be confirmed by forming a topological Josephson junction (JJ) where the supercurrent is carried by a single electron instead of a Cooper pair resulting in a $4\pi$ periodic supercurrent proportional to $\sin(\phi/2)$ [15]. Here, $\phi$ denotes the gauge-invariant superconducting phase difference across the JJ. The topological JJ exhibits the fractional ac Josephson effect where the Shapiro voltage steps [16] are quantized at $V_Q = hf_{\text{mw}}/e$ unit under microwave irradiation with a microwave frequency of $f_{mw}$ [15, 17]. The fractional ac Josephson effect has been observed in JJs made of InSb nanowires [18], HgTe-based TIs [19], and Dirac semimetals [20]. However, numerous researchers have claimed that TI JJs exhibit the conventional Shapiro steps only [21-23] to raise controversial issues. The formation of a highly transparent superconducting contact is essential for observing the topological Josephson effect [24] because the superconducting proximity effect in the Josephson weak link is strongly dependent on the transparency at the contacting interface [25].

Here we report the fabrication and characterization of superconducting proximity junctions made of Sb-doped $Bi_2Se_3$ TI NRs. The axial MC curves exhibited AB and AAS oscillations owing to the topological surface states of the TI NR under the application of $B_{axial}$. We observed the sub-harmonic gap structures of differential conductance below the superconducting transition temperature ($T_c$ = 1.3 K) of Al electrodes because of multiple Andreev reflections (MARs) indicating the successful formation of highly transparent contacts between the superconducting Al electrodes and TI NR. The interface transparency was estimated to be $\tau$ = 0.83, which is among the highest values reported so far for the TI JJs [26, 27]. Temperature dependences of the critical current in TI NR JJs are consistent with the short and ballistic junction model, which confirms the formation of highly transparent superconducting contacts on the TI NR. Our observations suggest that the TI NR JJs would be a useful platform for studying the topological Josephson effect for developing novel quantum information devices.

## 2. Experiments

We synthesized single-crystalline Sb-doped $Bi_2Se_3$ NRs using the chemical vapor deposition method with Au nanoparticles as a catalyst [28, 29]. From energy-dispersive X-ray spectroscopy, we observed that the atomic percentages of Bi, Sb, and Se were approximately

36.0, 5.5, and 58.5 %, respectively. resulting in the dopant ratio $x$ for $(Bi_{1-x}Sb_x)_2Se_3$ was approximately equal to 0.13 [11]. The TI NRs used in our research had a rectangular cross section with thickness and width ranging from 80 to 170 nm and 260 to 480 nm, respectively. Moreover, the total length of the cross section was equal to several tens of micrometers. We mechanically transferred the individual NRs to an $n$-doped silicon substrate covered with a 290 nm-thick $SiO_2$ layer after the growth of TI NRs. The electrode patterns were defined using standard electron-beam lithography followed by electron-beam evaporation of Ti (5 nm)/Al (140 nm). We treated the surface of the TI NR with oxygen plasma and a 6:1 buffered oxide etch to remove the electron-beam resist residues and the native oxide layer after the resist development. Prior to the metal deposition, we irradiated the TI NR with $Ar^+$ ions accelerated with a beam voltage of 200 V for 30 s without breaking the vacuum in the evaporation chamber [30]. The scanning electron microscopy (SEM) image post the device fabrication process is shown in Fig. 1a. The channel lengths, $L_{ch}$, of the TI NR JJs ranged from 60 to 240 nm (see Table. 1). We measured the TI NR devices using a $^3$He refrigerator system (Cryogenic, Ltd.) with a base temperature of 0.3 K. Low-pass RC filters and π filters were connected in series with the measurement leads for low-noise measurement.

### 3. Results and discussion

The MC curve ($G$ vs. $B_{axial}$) exhibited oscillatory behavior superposed on a parabolic background when we applied an external magnetic field, $B_{axial}$, along the TI NR axis as illustrated in Fig. 1b. The width ($w$) and thickness ($t$) of the TI NR were equal to 472 and 88 nm, respectively. Moreover, the channel length ($L_{ch}$) was equal to 5.0 μm. The quasi-periodic oscillations of $\delta G$ are displayed as a function of $B_{axial}$ in Fig. 1c after subtracting the smooth background attributed to the semi-classical Kohler's rule [31]. The MC oscillations exhibited two different periods $\Delta B_1 = 133$ mT and $\Delta B_2 = 62$ mT that are dominant at higher and lower fields, respectively, than $B_{axial}$ = 0.2 T. We established the existence of two different oscillation periods though the fast Fourier transform (FFT) analysis of $\delta G(B_{axial})$ curve in Fig. 1d.

The magnetic flux period through the cross-sectional area of the TI NR was given by $\Delta\Phi_1$ = 1.05 $\Phi_0$ and $\Delta\Phi_2$ = 0.49 $\Phi_0$ corresponding to the $B_{axial}$ periods of $\Delta B_1$ and $\Delta B_2$, respectively. Here, we assumed that a 5-nm-thick native oxide layer existed on the surface of the TI NR [11]. The $h/e$ periodic oscillations occurred owing to AB oscillations caused by the surface states forming 1D subbands of TI NR [3, 5], whereas the $h/2e$-periodic oscillations occurred owing to AAS oscillations caused by the weak antilocalization effect along the circumference of the TI NR [32]. The AAS oscillations become insignificant at $B_{axial}$ fields greater than 0.2 T because of the suppression of coherent backscattering due to the breaking of time-reversal symmetry. The coexistence of the AB and AAS oscillations in a TI NR for $L_{ch} \gtrsim L_p$, where $L_p$ means the perimeter length of the NR, is consistent with our previous observations of adjustable quantum oscillations in TI NR [11]. This supports the existence of the topological surface states in TI NR.

The voltage *vs.* current $V(I)$ curves obtained from **JJ1** are shown in Fig. 2a. The $V(I)$ curve at $T$ = 1.3 K exhibits a linear Ohmic behavior indicating the normal-state resistance $R_n$ = 137 Ω, whereas the superconducting curve measured at $T$ = 0.3 K exhibits a dissipationless supercurrent. The inset image exhibits the hysteretic behavior of the critical current $(I_c)$ = 580 nA and the return current $(I_r)$ = 530 nA. The corresponding hysteresis parameter was obtained to be $\beta_c$ = 1.15 from the calculation results [17]. The excess current was given by $I_{exc}$ = 1.55 μA when the $V(I)$ curve in the high-bias region in the superconducting state was extrapolated to the zero $V$-axis (red dashed line) indicating the existence of Andreev reflection [33].

The Andreev reflection occurs at a highly transparent contact formed at the interface between a normal metal (N) and superconductor (S) where an incident electron in a normal metal is reflected as a hole, creating a Cooper pair in the superconductor [25]. In an SNS junction, the hole is reflected back as an electron at the other NS interface repeating the Andreev reflection process several times. This is called multiple Andreev reflections (MARs). The MARs can be evidenced by the conductance peaks at the subharmonic gap voltages, such as $V_n = 2\Delta_s/ne$, where $\Delta_s$ denotes the superconducting gap energy and $n$ is an integer [34-36]. Figure 2b demonstrates the differential conductance ($dI/dV$) vs. $V$ curves at different temperatures with subgap conductance peaks occurring at $V_n$ for $n = 1, 2, 3$, and 4. The plot of $nV_n$ as a function of $T$ is consistent with the Bardeen–Cooper–Schrieffer (BCS) theory of superconductivity [16], revealing $2\Delta_s/e = nV_n = 288$ μV at $T = 0.3$ K in Fig. 2c. Thus, we obtained $eI_{exc}R_n/\Delta_s = 1.47$ for **JJ1**. The interface transparency ($\tau$) estimated from the analytical calculation result of $I_{exc}$ [37] resulted in $\tau = 0.83$, which is one of the highest transparency values obtained for TI-based JJs [26, 27].

The subharmonic energy-gap structures obtained from other TI NR JJs are demonstrated in Fig. 2d. The temperature dependencies of the $dI/dV$ peak voltages shown in Fig. 2e and 2f are consistent with the BCS theory. Thus, we obtained $2\Delta_s/e = 176 – 288$ μV at $T = 0.3$ in this work (see Table 1). We observed that the $dI/dV$ peaks became rounded and their conductance enhancements were reduced for JJs with long $L_{ch}$, because of the diffusive motion of the electron-like (or hole-like) quasiparticles along the TI NR [34, 38]. However, the longest-channel device **JJ4** with $L_{ch} = 241$ nm still maintained highly transparent contacts at $\tau = 0.77$ (see Table 1) indicating stable formation of highly transparent superconducting contacts on the surface of the TI NRs in our research.

The progressive evolution of the $dI/dV$ subgap structures under the application of the

$B_{axial}$ field is shown in Fig. 3a and 3b. We observed that the superconducting gap energy of Al was suppressed by the external $B_{axial}$ field and disappeared at $B_{axial}$ = 100 mT. Since the MBS in TI NR are expected to occur at the $B_{axial}$ field corresponding to the half-flux quantum [3, 12], which is $\Phi = h/2e$, it is necessary to select a TI NR with a proper cross-sectional area to meet the $B$ field criterion or to choose a field-resilient superconducting electrodes with higher critical magnetic field than that of Al [24, 29].

Temperature-dependent $V(I)$ curves for **JJ1** are shown in Fig. 3c. The $I_c(T)$ data obtained for a threshold voltage ($V_{th}$) = 1 μV is demonstrated in Fig. 3d for the four different devices. We observed that the shapes of the $I_c(T)$ curves obtained from **JJ1** and **JJ2** were distinct from the exponentially decaying curves obtained from **JJ3** and **JJ5** that were typical for diffusive JJs in a long junction regime [35, 39]. The electronic mean free path and the Fermi velocity were given by $l_{mfp}$ = 51 nm and $v_F = 3 \times 10^5$ m/s, respectively [11], resulting in the diffusion coefficient ($D$) = 51 cm$^2$/s. Hence, the Thouless energy ($E_{Th}$) was given by $\hbar D/L_{ch}^2$ = 960 and 730 μeV, where $\hbar$ denotes the reduced Planck constant, for **JJ1** and **JJ2**, respectively. Thus, we inferred that **JJ1** and **JJ2** belonged to the short ($E_{Th} \gg \Delta_s$) [40] and ballistic ($L_{ch} \sim l_{mfp}$) JJ regime.

In the short and ballistic junction model [41], $I_c$ is given by the maximum value of the supercurrent $I_s$ expressed by [42]

$$I_s(\phi,T)R_n = \alpha \frac{\pi \Delta_s(T)}{2e} \frac{\sin(\phi)}{\sqrt{1-\tau\sin^2(\phi/2)}} \tanh\left(\frac{\Delta_s(T)}{2k_BT}\sqrt{1-\tau\sin^2(\phi/2)}\right) \quad (1),$$

where $\alpha$ is a fitting parameter. The black and red solid lines represent the best fits to Eq. (1) using $\alpha$ = 0.25 for **JJ1** and **JJ2**, respectively. The superconducting coherence length, defined as $\xi = \hbar v_F/\Delta_s$ for a ballistic JJ [16], was found to be $\xi$ = 1.37 μm for both JJs. This was approximately twenty-times longer than $L_{ch}$ and satisfied the short-junction condition ($L_{ch} \ll \xi$) [40].

The Thouless energies of other devices are given by $E_{Th}$ = 87 and 68 μeV for **JJ3** and **JJ5**, respectively. These values were smaller than the value of $\Delta_s$ (106 and 88 μeV for **JJ3** and **JJ5**, respectively) obtained from the subgap structures due to MARs. Therefore, we inferred that **JJ3** and **JJ5** belonged to the long ($E_{Th} < \Delta_s$) and diffusive ($L_{ch} \gg l_{mfp}$) junction regime. $I_c$ was given by $eI_cR_n = aE_{Th}[1-b\exp(-aE_{Th}/3.2k_BT)]$, where $a$ and $b$ were fitting parameters [40]. Figure 3d illustrates that the $I_c(T)$ data of **JJ3** and **JJ5** fit well with the theoretical model using $a$ = 1.0 and 0.9 as well as $b$ = 1.4 and 1.3, respectively, which were comparable to the ones from other nano-hybrid JJs [35]. The superconducting coherence length given by $\xi = (\hbar D/\Delta_s)^{1/2}$ for a diffusive JJ [40] was equal to 180 and 200 nm for **JJ3** and **JJ5**, respectively. These values satisfied the long-junction condition ($L_{ch} \gg \xi$) [40].

We observed an anomalous dip feature in the $I_c(T)$ data near $T$ = 0.4 K for **JJ1**, **JJ2**, and **JJ3**, accompanied by an intermediate voltage jump in the $V(I)$ curve, as can be seen in Fig. 3c. The intermediate voltage jump was reported in other TI NR JJs, which was attributed to the phase slip due to local inhomogeneities in the TI NR [43]. However, the $I_c(T)$ curve exhibited monotonously increasing trend corresponding to a decrease in temperature, which was contrary to our observations. Moreover, an abrupt increase of $I_c$ at low temperatures below 0.2 $T_c$ was also reported in other TI NR JJs [44], which was ascribed to the additional supercurrent due to the Andreev bound states around the circumference of the TI NR. However, we could not study the $I_c$-upturn phenomenon because of the limitation of the base temperature (approximately 0.3 K) in our apparatus.

We successfully formed highly transparent superconducting contacts on TI NRs and observed supercurrent in the TI NR JJs. The AB and AAS oscillations of the axial magnetoconductance of the TI NR indicated the existence of topological surface states along the circumference of the NR. The subharmonic gap structures of the differential conductance due to the MARs were consistent with theoretical explanations revealing the superconducting gap energy. The value of the transparency of the superconducting contacts ($\tau = 0.83$) that was calculated from excess current in our research was among the highest values reported for TI JJs yet. Our research can be useful for future studies on the fractional AC Josephson effects in TI NR JJs due to MBS.

**Declaration of competing interest**

The authors declare that they have no known competing financial interests or personal relationships that could have influenced the work reported in this paper.

**Acknowledgment**

This study was supported by the NRF of Korea through the Basic Science Research Program (2018R1A3B1052827) and the SRC Center for Quantum Coherence in Condensed Matter (2016R1A5A1008184). The work at UC Davis was supported by the U.S. National Science Foundation (Grant No. DMR-2105161).

**Figure Captions**

**Figure 1.** (a) Representative SEM image of the TI NR JJs. The voltage drop was measured between V+ and V− when the current was biased from I+ to I−. (b) Axial MC, $G(B_{axial})$, curve of TI NR measured at $T = 2.0$ K. Inset: (Upper) SEM image of the TI NR used for MC measurement. Scale bar is equal to 2 μm. (Lower) Schematic view of the TI NR device. The $B_{axial}$ field was applied along the NR axis. (c) Differential MC curve as a function of $B_{axial}$ field. $\delta G$ was obtained after subtracting the smooth background from the MC curve in (b). The dotted lines are drawn with an interval of $\Delta B_1 = 133$ mT. (d) FFT amplitudes of $\delta G(B_{axial})$ curve in (c). Two arrows indicate the $h/e$- and $h/2e$-periodic oscillations, respectively. Inset: Height profile of TI NR obtained using atomic force microscopy.

**Figure 2.** (a) $V(I)$ curves of **JJ1** measured at $T = 0.3$ and 1.3 K, respectively. Inset: Magnified view near the switching current region. $I_c$ and $I_r$ denote the critical and return current, respectively, whereas the red and blue arrows denote the corresponding voltage jump and drop, respectively. (b) Temperature dependence of the differential conductance $dI/dV$ vs. $V$ curve. The arrows indicate the $dI/dV$ peaks with an integer index $n$. $dI/dV(V)$ curves were offset vertically for clarity. (c) Temperature-dependent $nV_n$ data (symbols) extracted from (b) and $\Delta_s(T)$ based on the BCS theory (solid line). (d) Normalized $dI/dV(V)$ curves obtained from four distinct devices measured at $T = 0.3$ K. $dI/dV$ and $V$ axes were normalized by the normal-state conductance $G_N$ and $V_1 = 2\Delta_s/e$, respectively, for each device. $dI/dV(V)$ curves were offset vertically for clarity. (e–f) Temperature-dependent $nV_n$ data (symbols) and theoretical values (solid line) of $\Delta_s(T)$ for **JJ2** and **JJ3**, respectively.

**Figure 3.** (a) Color scale plot of d$V$/d$I$ of **JJ1** as a function of $V$ and $B_{axial}$ at $T$ = 0.3 K. (b) $B_{axial}$ field-dependent d$V$/d$I(V)$ curves that were offset vertically for clarity. (c) $T$-dependent $V(I)$ curves (symbols). The solid lines are guides to the eye. The current bias was swept from negative to positive polarity. (d) $I_c(T)$ curves (symbols) were obtained from four distinct JJs. The solid lines are theoretical calculations (see text).

**Table 1.** Geometrical dimensions and physical parameters of TI NR JJs.

| Device | $L_{ch}$ (nm) | $w$ (nm) | $t$ (nm) | $I_c$ (nA) | $I_{exc}$ (μA) | $R_n$ (Ω) | $I_c R_n$ (μeV) | $\Delta_s$ (μeV) | $\tau$ |
|---|---|---|---|---|---|---|---|---|---|
| **JJ1** | 59  | 265 | 92  | 580 | 1.55 | 137 | 79.5 | 144 | 0.83 |
| **JJ2** | 68  | 264 | 92  | 490 | 1.40 | 150 | 73.5 | 144 | 0.83 |
| **JJ3** | 196 | 303 | 166 | 207 | 1.05 | 129 | 26.7 | 106 | 0.79 |
| **JJ4** | 241 | 300 | 166 | 90  | 0.75 | 146 | 13.1 | 93  | 0.77 |
| **JJ5** | 223 | 472 | 88  | 97  | 0.37 | 223 | 21.6 | 88  | 0.71 |

**Figure 1.**

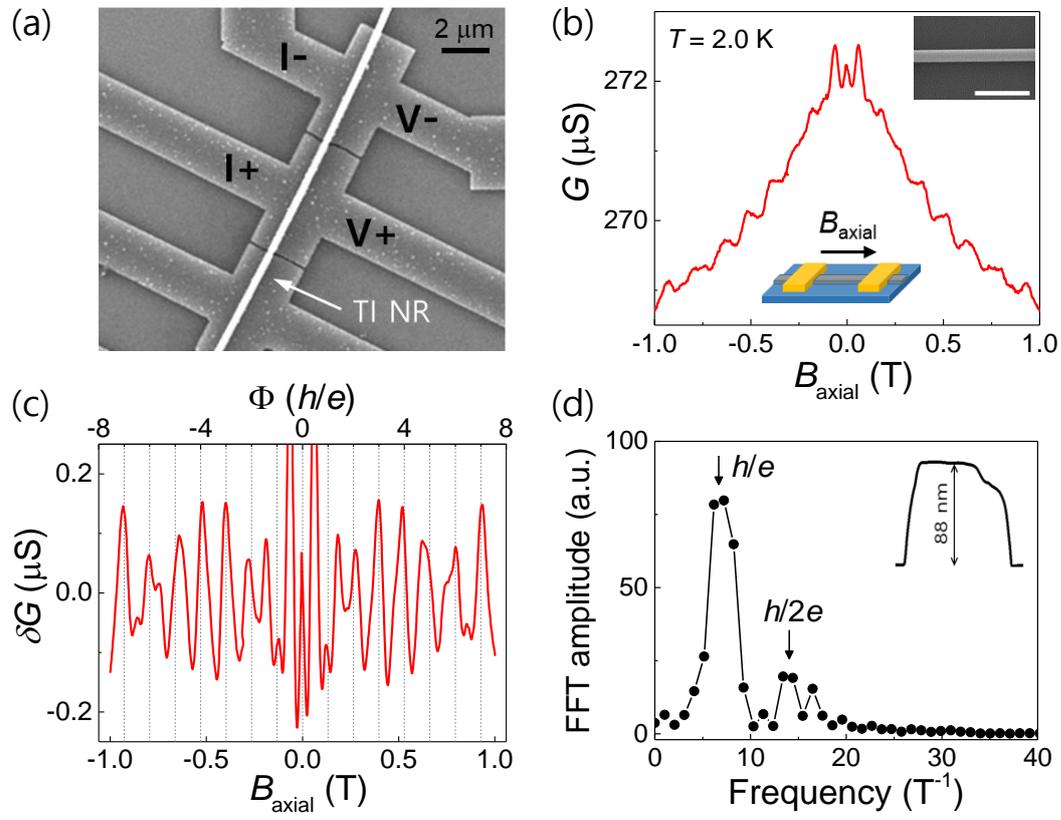

**Figure 2.**

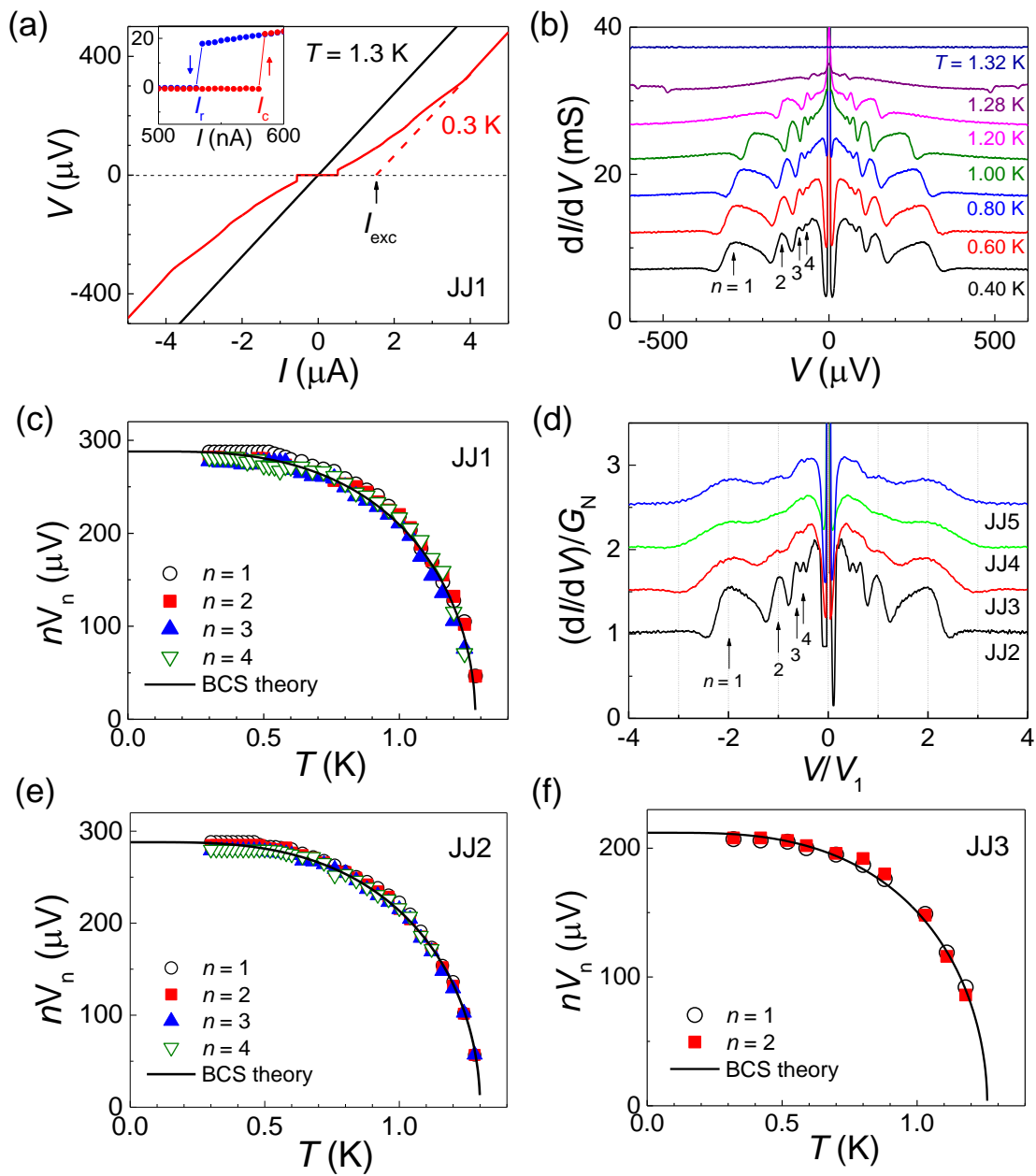

**Figure 3.**

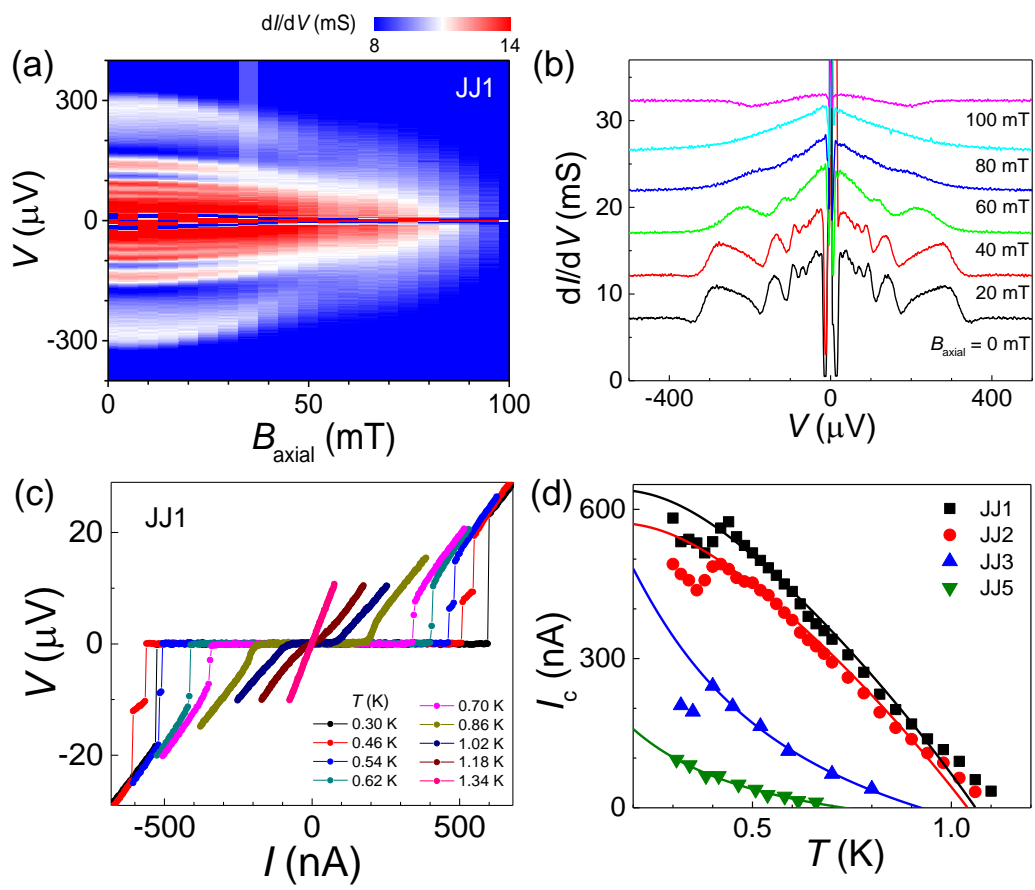